\documentclass[10pt]{npqcd}
\begin{document}
\renewcommand\nextpg{\pageref{pgs1}}\renewcommand\titleA{
The critical exponents of the QCD (tri)critical endpoint within exactly solvable models
}\renewcommand\authorA{
 A. I. Ivanytskyi${}^{\mathrm{a}}$,
 K. A. Bugaev${}^{\mathrm{b}}$
}\renewcommand\email{
 e-mail:\space \eml{a}{a\_iv\_@ukr.net}, \eml{b}{bugaev@th.physik.uni-frankfurt.de}\\[1mm]
}\renewcommand\titleH{
 The critical exponents of the QCD (tri)critical endpoint within exactly solvable models
}\renewcommand\authorH{
 A. I. Ivanytskyi, K. A. Bugaev
}\renewcommand\titleC{\titleA}\renewcommand\authorC{\authorH}\renewcommand\institution{
 Bogolyubov Institute for Theoretical Physics, Kiev, Ukraine


}\renewcommand\abstractE{
The critical indices $\alpha'$, $\beta$, $\gamma'$ and $\delta$ of the Quark Gluon Bags with Surface Tension  Model with the tricritical and critical endpoint are calculated as functions of the usual  parameters of this model  and two newly introduced parameters (indices). The critical indices are compared with that ones of other models. The universality class of the present model with respect to values of the model parameters is discussed. The scaling relations for the found critical exponents  are verified and  it is demonstrated  that for the standard definition of the index $\alpha'$ some of them are not  fulfilled in general case. Although it is shown that the specially defined index $\alpha'_s$ recovers the scaling relations, another possibility, an existence of the non-Fisher universality classes,  is also discussed.
}


\begin{article}
\section{Introduction}

Investigation of the properties of strongly interacting matter equation of state has become a focal point of modern
nuclear physics of high energies. The low energy scan programs performed nowadays at CERN SPS and BNL RHIC
are aimed at the discovery of the (tri)critical endpoint of the quantum chromodynamics (QCD) phase diagram.
Despite many theoretical efforts neither an exact location nor the properties of the QCD (tri)critical endpoint are well known \cite{Ivanytskyi:Shuryak:sQGP}.  Therefore, the thorough theoretical  investigation of the QCD endpoint properties  are required in order  to clarify  whether this  endpoint is critical or  tricritical.

This work is devoted to calculation of the critical exponents of the QCD endpoint of both critical (CEP) and tricritical (triCEP) types. Unfortunately, such a task cannot be solved within the QCD itself. Moreover, even the possibilities of the lattice QCD are nowadays  very limited in this respect. Then,  unavoidably, one has to use some models of the (tri)CEP.
The most popular models of this kind  are the quark-meson  model \cite{Ivanytskyi:QMM:1,Ivanytskyi:QMM:2}
and  the  extended Nambu--Jona-Lazinio model \cite{Ivanytskyi:PNJL:1}. Despite a great popularity both of these models contain
a significant weakness, they are mean-field ones and hence there is no reason to expect that their critical exponents
could differ from that ones of the Van der Waals model  equation of state \cite{Ivanytskyi:Stanley:71}.
Therefore, to study the properties of non-classical endpoints one has to investigate the non-mean-field models.

The gas of bags model \cite{Ivanytskyi:GasOfBags:81} is an example of such a non-mean-field model. Despite a significant success of this model in describing of the deconfinement  phase transition, it can not  generate the (tri)CEP in a natural way.
The problem was solved in the Quark Gluon Bags with Surface Tension Model  (QGBSTM)
\cite{Ivanytskyi:Bugaev_07_physrev, Ivanytskyi:Bugaev_07_physpartnucl}. This exactly solvable model accounts for the surface effects which play a decisive role in the critical phenomena and employes the same mechanism of the (tri)CEP generation which is typical for the liquid-gas phase transition (PT) and which is also used in the Fisher droplet model (FDM) \cite{Ivanytskyi:Fisher_67} and in the statistical multifragmentation model (SMM) \cite{Ivanytskyi:Bondorf,Ivanytskyi:Bugaev_00}: the endpoint of the 1-st order PT appears due to vanishing of the surface tension coefficient at this point which leads to the indistinguishability between the liquid and gas phases. However, the surface tension coefficient in the QGBSTM has the region of negative values, which is a principally different feature of this model  compared  to the FDM, SMM
and all other statistical  models of the liquid-gas PT.
Note  that just this feature  provides an existence of the cross-over at small values of the baryonic chemical potential $\mu$ in the QGBSTM with triCEP  \cite{Ivanytskyi:Bugaev_07_physrev}
and with CEP \cite{Ivanytskyi:Bugaev_00}, and also it generates an  additional PT in the model with triCEP at large values of  $\mu$.
Therefore, it is very important and  interesting  to
study the critical indices of such a novel statistical model as  the QGBSTM,  to determine its class of universality and to examine  how the latter is related to  that  ones of the FDM and SMM.

The paper is organized as follows. A description of the QGBSTM is given in Section \ref{Ivanytskyi:model}. In Section
\ref{Ivanytskyi:critexp} the critical exponents of the CEP and triCEP are calculated. This section is also devoted to the
analysis of scaling relations between the found critical exponents. Conclusions are given in Section \ref{Ivanytskyi:conclusions}.


\section{Quark Gluon Bags with Surface Tension Model}
\label{Ivanytskyi:model}

An exact solution of the QGBSTM was found in \cite{Ivanytskyi:Bugaev_07_physrev}. The relevant degrees of freedom in this model are the quark gluon plasma (QGP) bags and hadrons. The attraction between them is accounted like in the original statistical bootstrap model \cite{Ivanytskyi:SBM} via  many  sorts of the constituents, while the repulsion between them is introduced a la Van der Waals equation of state \cite{Ivanytskyi:GasOfBags:81,Ivanytskyi:Bugaev_07_physrev}. An essential element of the QGBSTM  is  the $T$ and $\mu$ dependence of its  surface tension coefficient $T \Sigma(T,\mu)$ (here $\Sigma(T,\mu)$ is the reduced surface tension coefficient). Let us denote the nil line of the reduced surface tension coefficient in the $T-\mu$ plane as  $T_\Sigma(\mu)$, i.e. $\Sigma(T_\Sigma,\mu)=0$. Note that for a given $\mu$ the surface tension is negative (positive) for $T$ above (below) $T_\Sigma(\mu)$ line. Here it is appropriate to say a few words about the negative values of $\Sigma$ which is a distinctive feature of QGBSTM compared to other models. There is nothing wrong or unphysical with the negative values of surface tension coefficient, since $T \Sigma\, v^\kappa$ is the surface  free energy of the bag of mean volume $v$ and, hence, as any free energy,  it contains the energy part $e_{surf}$ and  the entropy part $s_{surf}$ multiplied by temperature $T$ \cite{Ivanytskyi:Fisher_67}. Therefore, at low temperatures the energy part dominates and surface  free energy is positive, whereas at high temperatures the number of bag configurations with large surface  drastically increases  and it exceeds  the  Boltzmann suppression and, hence, the surface free energy becomes negative since $s_{surf} > \frac{e_{surf}}{T}$. Such a behavior of the surface free energy  can be derived within the exactly solvable model of surface deformations known as Hills and Dales Model \cite{Ivanytskyi:Bugaev_05_Physrev}.  

In the grand canonical ensemble the pressure of   QGP and hadronic phase are, respectively,  given by
\begin{eqnarray}
\label{Ivanytskyi:pQ}
p_Q(T,\mu) & = & Ts_Q(T,\mu), \\
\label{Ivanytskyi:pH}
p_H(T,\mu)& = & Ts_H(T,\mu) = T\left[F_H(s_H,T,\mu)+u(T,\mu)I_\tau(\Delta s,\Sigma)\right],\\
\label{Ivanytskyi:FH}
F_H(s_H,T,\mu) & = & \sum_{j=1}^n g_j e^{\frac{b_j \mu}{T} -v_js_H} \phi(T,m_j)\quad{\rm and}\quad
%
%
I_\tau(\Delta s,\Sigma)  =  \int\limits_{V_0}^{\infty}\frac{dv}{v^\tau}e^{-\Delta s v-\Sigma v^{\kappa}},
\end{eqnarray}
%
where  $\Delta s\equiv s_H (T,\mu)-s_Q(T,\mu)$. The particle density of a  hadron of mass $m_j$, baryonic charge $b_j$, eigenvolume $v_j$ and degeneracy $g_j$ is denoted as
$\phi_j(T,m_j)\equiv \frac{1}{2\pi^2}\int\limits_0^{\infty}\hspace*{-0.1cm}p^2dp\, e^{\textstyle-\frac{(p^2~+~m_j^2)^{1/2}}{T}}$.
Here, as in \cite{Ivanytskyi:Bugaev_07_physrev}, it is assumed  that the functions  $u(T,\mu)$ and $s_Q(T,\mu)$, which are the parameters of the present model, and   their first and second derivatives with respect to  $T$ and $\mu$   are finite  everywhere at the $T-\mu$ plane. In the continuous part of the spectrum of bags denoted as $uI_\tau(\Delta s,\Sigma)$ the surface of a QGP bag of volume $v$ is parameterized by the term  $v^\kappa$. Usually one chooses $\kappa=\frac{2}{3}$ in 3-dimensional case (or $\kappa=\frac{d-1}{d}$ for the  dimension $d$), but in what follows it  is  regarded as free parameter of  the range $0<\kappa<1$.

\begin{figure}
\centering
  \includegraphics[bb= 300 0 850 350,width=.4\textwidth]{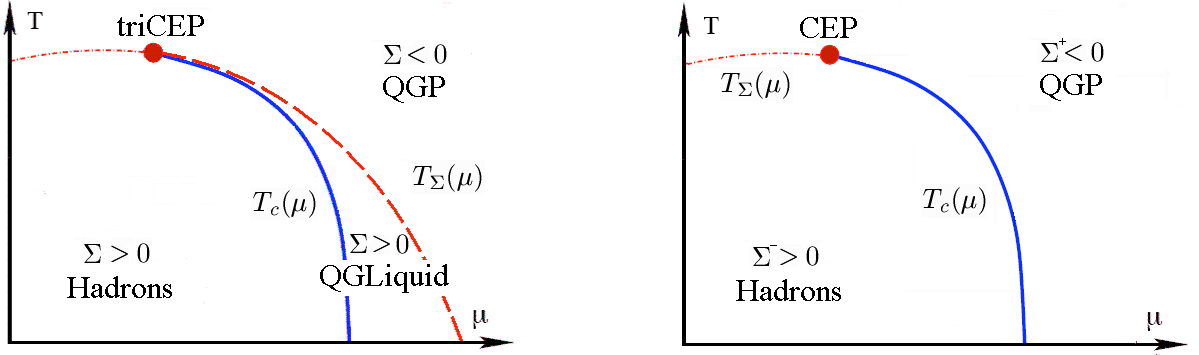}\\
  \caption{
A schematic  phase diagrams with triCEP (left panel) and CEP (right panel) in the $\mu-T$ plane. The dashed  curve indicates the nil line of the  surface tension coefficient  $T_\Sigma(\mu)$, below (above) which $\Sigma$  is positive (negative). The  deconfinement PT line  $T_c(\mu)$ is shown by the full curve  for $\mu > \mu_{cep}$. At the phase diagram with CEP $T_c(\mu)$ and $T_\Sigma(\mu)$ lines coincide exactly for $\mu > \mu_{cep}$. A cross-over (the short dashed  curve) takes place along the line $T_\Sigma(\mu)$  for $\mu \le \mu_{cep}$.  The  cross-over and PT regions are separated by (tri)CEP (filled  circle).
  }\label{Ivanytskyi:Figure}
\end{figure}

The system pressure, that corresponds to a dominant  phase, is given by the largest value between $p_Q$ and $p_H$ for each set of $T$ and $\mu$. As usual,  the deconfinement PT occurs when pressure of  QGP  gets equal to that one of the hadron gas, which is nothing else as the Gibbs criterion. The sketch of $T-\mu $ phase diagram with triCEP is shown in the left panel of Fig. \ref{Ivanytskyi:Figure}. Suppose that the necessary conditions for the deconfinement PT outlined in
\cite{Ivanytskyi:Bugaev_07_physrev} are satisfied and its  transition temperature is given by the function $T_c(\mu)$ for $\mu\ge\mu_{cep}$. The necessary condition of the triCEP occurrence is that at the nil line of the surface tension coefficient there exists the surface tension induced PT of 2-nd (or higher order)  \cite{Ivanytskyi:Bugaev_07_physrev} for $\mu\ge\mu_{cep}$  and $T_c(\mu)\le T_\Sigma(\mu)$ for these $\mu$ values. The both of these  inequalities become equalities only at triCEP. Moreover, at  triCEP the phase coexistence curve $T_c(\mu)$  is a tangent (not intersecting!) line to the nil line of the surface tension coefficient $T_\Sigma(\mu)$. For  $\mu <  \mu_{cep}$ the deconfinement PT degenerates into a cross-over since in this region $\Sigma < 0$ \cite{Ivanytskyi:Bugaev_07_physrev} and, hence, the system pressure is defined  by a solution of  Eq. (\ref{Ivanytskyi:pH}). It is necessary to stress here that QGP exists as real quark gluon liquid (QGLiquid) with the pressure (\ref{Ivanytskyi:pQ}) only for $T_c(\mu)\le T\le T_\Sigma(\mu)$ and $\mu\ge\mu_{cep}$, outside this region the system is defined by Eq. (\ref{Ivanytskyi:pH}). Now let us discuss phase diagram with CEP (see the right panel in Fig. \ref{Ivanytskyi:Figure}). The only difference with the case of triCEP is that $T_c(\mu)$ and $T_\Sigma(\mu)$ lines coincide exactly for $\mu\ge\mu_{cep}$. Therefore, QGLiquid exists only at the PT line.

An  actual parameterization of the reduced  surface tension coefficient $\Sigma(T,\mu)$ is taken from \cite{Ivanytskyi:Bugaev_07_physrev,Ivanytskyi:Bugaev_09}:
\begin{eqnarray}
\label{Ivanytskyi:Sigma}
\Sigma(T,\mu)&=&\frac{\sigma_0}{T}\cdot\left| \frac{T_\Sigma(\mu)-T}{T_\Sigma(\mu)}\right|^\zeta
{\rm sign} \left(  T_\Sigma(\mu)-T \right) \quad {\rm for ~ triCEP} \,,\\
\Sigma^\pm(T,\mu)&=&\mp\frac{\sigma_0}{T}\cdot
\left( T_{cep}-T+\frac{dT_c}{d\mu}(\mu_{cep}-\mu)\right)^{\xi^\pm}
\left| \frac{T_\Sigma(\mu)-T}{T_\Sigma(\mu)}\right|^{\zeta^\pm}\quad {\rm for ~ CEP} \,.
\end{eqnarray}
In what follows the coefficient $\sigma_0$ is assumed to be a positive constant, i.e.  $\sigma_0={\rm const}>0$,  but  it is easy to show that   the obtained results hold, if  $\sigma_0 >0$ is  a smooth function of $T$ and $\mu$.


In the vicinity of (tri)CEP the behavior of both the deconfinement PT curve and the nil  surface tension coefficient line in the $\mu-T$ plane is parameterized via a single  parameter $\xi^T>0$:
\begin{equation}
\label{Ivanytskyi:xiT}
T_{cep}-T_\Sigma(\mu)\sim (\mu-\mu_{cep})^{\xi^T} \quad{\rm and}\quad T_{cep}-T_c(\mu)\sim(\mu-\mu_{cep})^{\xi^T} \,,
\end{equation}
since, as discussed above,   $T_c(\mu)$ and $T_\Sigma(\mu)$ lines are tangent to each other  at the endpoint in the model with triCEP and coincide in the model with CEP. This is a new index which was not considered both in the FDM and SMM and as we show below, it is responsible for a new universality class compared to other exactly solvable models.

Since the entropy density and the baryonic density are, respectively,  defined as $T$ and $\mu$ partial derivatives of the corresponding pressure, then using (\ref{Ivanytskyi:pQ}) and (\ref{Ivanytskyi:pH}) the Clapeyron-Clausius equation $\frac{d\mu_c}{dT}=-\frac{S_H-S_Q}{\rho_H-\rho_Q}\Bigl|_{T=T_c}$ for the model with triCEP can be explicitly  rewritten as
\begin{equation}\label{Ivanytskyi:EqVII}
\frac{d\mu_c}{dT}=-\frac{A_T-\frac{\partial\Sigma}{\partial T}uI_{\tau-\kappa}(0,\Sigma)}
{A_\mu-\frac{\partial\Sigma}{\partial\mu}uI_{\tau-\kappa}(0,\Sigma)}\biggl|_{T=T_c} \,,
\end{equation}
where the following notation is used
$A_i\equiv\frac{\partial F_H}{\partial i}+\frac{\partial u}{\partial i}I_\tau+\frac{\partial s_Q}{\partial i}
\left(\frac{\partial F_H}{\partial s}-1\right)$ for $i\in\{T,\mu\}$. Note that for the model with CEP the term
$\frac{\partial\Sigma}{\partial i}uI_{\tau-\kappa}(0,\Sigma)$ does not appear in (\ref{Ivanytskyi:EqVII}).
Let us   parameterize  the  behavior of  the numerator and denominator in
(\ref{Ivanytskyi:EqVII}) at  the (tri)CEP vicinity as
\begin{equation}
A_T\bigl|_{T=T_c}  \sim  (T_{cep}-T_c(\mu))^{\chi+\frac{1}{\xi^T}-1}\quad{\rm and}\quad
A_\mu\bigl|_{T=T_c}\sim  (T_{cep}-T_c(\mu) )^\chi,
\end{equation}
where  $\chi\ge\max(0,1-\frac{1}{\xi^T})$ denotes  another  new index. The latter inequality  follows from the fact that the integral $I_\tau$ and functions $F_H$, $u$, $s_Q$ together  with their derivatives are finite for any finite values of  $T$ and $\mu$. An introduction of the index $\chi$ is quite general and could be done for any model with the PT of the liquid-gas type, since it unavoidably  appears  from the Clapeyron-Clausius equation, which  is a direct  consequence of the  Gibbs criterion for  phase equilibrium. To our best knowledge, this index was never used for the calculation of critical exponents.




\section{Critical exponents of the QGBSTM}
\label{Ivanytskyi:critexp}

The standard set of critical exponents  $\alpha',\beta$ and $\gamma$
\cite{Ivanytskyi:Stanley:71,Ivanytskyi:Fisher_70} describes the $T$-dependence of the system near (tri)CEP:
\vspace*{-0.3cm}
\begin{eqnarray}
\label{Ivanytskyi:alphadef}
C_\rho  & \sim\ &
|t|^{-\alpha'},  \hspace*{0.1cm}
{\rm for}  \quad  t \le 0\quad{\rm and}\quad\rho=\rho_{cep},\\
\Delta\rho & \sim  & |t|^\beta,  \hspace*{0.4cm}
{\rm for}  \quad t \le 0, \\
\Delta K_T & \sim  & |t|^{-\gamma'},  \hspace*{0.1cm}
{\rm for}  \quad t \le  0,
\end{eqnarray}
where $\Delta\rho\equiv(\rho_Q-\rho_H)_{T=T_c}$ defines the order parameter, $C_\rho\equiv\frac{T}{\rho}(\frac{\partial S}{\partial T})_\rho$ denotes the specific heat at the critical density and $\Delta K_T\equiv(K_T^H-K_T^Q)_{T=T_c}$ is the discontinuity in the isothermal compressibility $K_T\equiv\frac{1}{\rho}(\frac{\partial\rho}{\partial p})_T$ across the PT line, the variable $t$ is the  reduced temperature $t \equiv \frac{T-T_{cep}}{T_{cep}}$.
The   critical isotherm shape is given by the  index $\delta$ \cite{Ivanytskyi:Fisher_70,Ivanytskyi:Stanley:71}
(hereafter the tilde indicates that $T=T_{cep}$):
\begin{equation}
p_{cep}-\widetilde{p}\sim(\rho_{cep}-\widetilde{\rho})^\delta  \hspace*{0.1cm}
{\rm for}  \quad t = 0.
\end{equation}

In this work we want to present the results of our calculations which are exhaustively  explained in  \cite{Ivanytskyi:Ivanytskyi}. Not going into details here we give the standard set of critical exponents (see Table \ref{Ivanytskyi:table1}) and discuss the related physics issues. Since in some aspects the QGBSTM is similar to the FDM and SMM it is interesting to compare its critical exponents  with that ones of the FDM \cite{Ivanytskyi:Fisher_67} and  SMM \cite{Ivanytskyi:Reuter}. The QGBSTM with CEP generates quite unique set of the critical exponents which are independent of $\tau$. Therefore, this model belongs to different universality class, than that one of the SMM and FDM. As it is seen from  Table \ref{Ivanytskyi:table1}, QGBSTM with triCEP has two regimes switched by parameter $\chi$. Since  the FDM and SMM implicitly treat the parameter $\chi=0$,  then it is most  natural to compare their  critical exponents with the QGBSTM results just for this case. In the QGBSTM with triCEP there is a regime, when its index  $\delta |_{\chi=0}=\frac{\tau-1}{2-\tau}$ matches the SMM result \cite{Ivanytskyi:Reuter}. Moreover,   it is easy to see that in this regime  all other indices the QGBSTM and SMM coincide for $\xi^T \le 1$.  The spectrum of values of  the  QGBSTM critical indices is  more rich than the corresponding spectra   of the FDM and SMM since this model contains two new indices $\xi^T$ and $\chi$.
\begin{table}[t]
%
%
\hspace*{2.5cm}
\begin{tabular}{|c|c|c|c|}
\hline
%
%
%
        & \multicolumn{2}{c|}{ Model with triCEP} &   \multicolumn{1}{c|}{ Model with CEP} \\  \hline

            &\hspace*{1.5cm} $\chi=0$\hspace*{1.5cm}&\hspace*{1.5cm}$\chi>0$ \hspace*{1.5cm}& \\ \hline
   $\alpha'$ & \multicolumn{2}{c|}{  $2-2\min(1,\frac{1}{\xi^T})$}  &  $2-2\min(1,\frac{1}{\xi^T})$  \\ \hline
  $\beta$   & \multicolumn{2}{c|}{ $\frac{\zeta}{\kappa}(2-\tau)+\min\left[\chi,\frac{\zeta}{\kappa}\min(\kappa,\tau-1)-\frac{1}{\xi^T}\right]$} &
               $\min(\beta^+,\beta^-)$ \\ \hline
  $\gamma'$ & \multicolumn{2}{c|}{ $\frac{\zeta}{\kappa}-2\beta$}  & $\frac{1}{\xi^T}-\beta$ \\ \hline
  $\delta$  &
                               $\min^{-1}\left[\frac{\xi^T\zeta}{\max(\tau-1,\kappa)}-1,\frac{2-\tau}{\tau-1}\right]$&
                    \hspace*{0.2cm}$\left(\frac{\xi^T\zeta}{\max(\tau-1,\kappa)}-1\right)^{-1}$  &
                               $\frac{1}{\xi^T\beta^+}$   \\ \hline
  $\alpha'_s$ &  \multicolumn{2}{c|}{$2-\min(2,\frac{1}{\xi^T})-\beta$}  & $2-\min(2,\frac{1}{\xi^T})-\beta$ \\ \hline
\end{tabular}

%
%
\caption{\hspace*{-0.2cm}Critical exponents of QGBSTM. For the model with CEP the notation $\beta^\pm\hspace*{-0.1cm}\equiv\zeta^\pm\hspace*{-0.1cm}+\xi^\pm\hspace*{-0.1cm}-\hspace*{-0.1cm}\frac{1}{\xi^T}$ is introduced. The index $\alpha'_s$ describes the specific heat  difference
$\Delta C=(C_{\rho_H}-C_{\rho_Q})_{T=T_c}\sim |t|^{-\alpha'_s}$ for two phases \cite{Ivanytskyi:Fisher_70}.}
\label{Ivanytskyi:table1}
\vspace*{-0.25cm}
\end{table}

The explicit expressions for the QGBSTM critical exponents allow us to examine scaling relations between them. For the model with triCEP Fisher inequality ($\alpha'+2\beta+\gamma'\ge2$) and Griffiths inequality ($\alpha'+\beta(1+\delta)\ge2$) are not fulfilled in a general case, whereas the Liberman inequality ($\gamma'+\beta(1-\delta)\ge0$) is always obeyed \cite{Ivanytskyi:Ivanytskyi}. For the model with CEP the situation is similar. In order to 'save' the scaling inequalities it was suggested to replace the index $\alpha'$ by $\alpha'_s$ \cite{Ivanytskyi:Fisher_70}. This hypothesis resolves the problem of scaling inequalities only for the triCEP model. Therefore, it is quite possible that the non-Fisher universality classes exist.


\section{Conclusions}
\label{Ivanytskyi:conclusions}

The critical indices of the QGBSTM are found in terms of the model parameters. Two parameters $\xi^T>0$ and $\chi\ge\max(0,1-\frac{1}{\xi^T})$ are newly introduced and consequently the spectrum of the values of critical exponents of the present model is more rich compared to other models. It is shown there is a regime when QGBSTM with triCEP reproduces the critical exponents of the SMM with triCEP, whereas the critical exponents of the FDM are never reproduced by QGBSTM. Thus, for $\chi=0$ QGBSTM and SMM fall into the same universality class.

The direct calculations show that  for the standard definition of the critical index $\alpha'$ (found along  the critical isochore)  the Fisher and Griffiths scaling  inequalities are not always fulfilled, whereas the Liberman inequality is obeyed for any values of the model parameters. According to Fisher hypothesis the scaling relations for the index $\alpha'_s$ are verified. The possibility of the non-Fisher universality classes existence is discussed.

\end{article}
\label{pgs1}
\end{document}